\documentclass[sigchi]{acmart}

\def\BibTeX{{\rm B\kern-.05em{\sc i\kern-.025em b}\kern-.08emT\kern-.1667em\lower.7ex\hbox{E}\kern-.125emX}}

\newcommand{\matfff}[1]{\mathcal{M}^{\text{#1}}}
\newcommand{\dataset}[1]{\mathcal{D}^{\text{#1}}}

\usepackage{hyperref} 

\renewcommand\footnotetextcopyrightpermission[1]{} 
\begin{document}

\title{Hybrid Approaches to Detect Comments Violating Macro Norms on Reddit}

\author{Eshwar Chandrasekharan}
\affiliation{%
  \institution{Georgia Institute of Technology}
  \city{Atlanta}
  \state{Georgia}
  \country{USA}
}
\email{eshwar3@gatech.edu}

\author{Eric Gilbert}
\affiliation{%
  \institution{University of Michigan}
  \city{Ann Arbor}
  \state{Michigan}
  \country{USA}
}
\email{eegg@umich.edu}

\renewcommand{\shortauthors}{Eshwar Chandrasekharan and Eric Gilbert}

\begin{abstract}
In this dataset paper, we present a three-stage process to collect Reddit comments that are removed comments by moderators of several subreddits, for violating subreddit rules and guidelines. Other than the fact that these comments were flagged by moderators for violating community norms, we do not have any other information regarding the nature of the violations.
Through this procedure, we collect over 2M comments removed by moderators of 100 different Reddit communities, and \href{https://doi.org/10.5281/zenodo.3338698}{publicly release the data}.
Working with this dataset of removed comments, we identify 8 \textit{macro norms}---norms that are widely enforced on most parts of Reddit.
We extract these macro norms by employing a hybrid approach---classification, topic modeling, and open-coding---on comments identified to be norm violations within at least 85 out of the 100 study subreddits.
Finally, we label over 40K Reddit comments removed by moderators according to the specific type of macro norm being violated, and make this dataset \href{https://doi.org/10.5281/zenodo.3338698}{publicly available}.
By breaking down a collection of removed comments into more granular types of macro norm violation, our dataset can be used to train more nuanced machine learning classifiers for online moderation.
\end{abstract}


\maketitle

\section{Introduction}
\noindent Reddit is organized into over one million user-created online communities\footnote{\url{http://redditmetrics.com/history}} known as \textit{subreddits}, and the platform has a multi-layered architecture for regulating behaviors. At a site-wide level, Reddit has content and anti-harassment policies that all subreddits are expected to follow. Whenever subreddits are found to continuously violate policies, Reddit is known to intervene and ban these forums and associated user accounts~\cite{Chandrasekharan2017you}. 
Additionally, each subreddit has its own set of subreddit-specific rules and guidelines regarding submissions, comments, and user behaviors~\cite{casey2018reddit,kasunic2018least}. Like many online platforms, Reddit relies on volunteer\textit{ moderators} (or ``mods'') to enforce the rules and guidelines within subreddits by sanctioning norm violating content (e.g., comment removal) and users (e.g., banning).

We compile a dataset of comments that moderators removed from Reddit for violating subreddit rules and guidelines. Other than the fact that these comments were flagged by the moderators as being violations, we do not have any other information regarding the nature of these comments. 
Our goal is to generate training data for building machine learning tools that can automatically detect nuanced norm violations, including antisocial behaviors like trolling~\cite{cheng2015antisocial} and hate speech~\cite{davidson2017automated,salminen2018anatomy, saha2019prevalence}, and thereby mimic the actions taken by moderators of different subreddits. 
In order to do this task, we propose an approach that is more granular than a simple binary classification task (i.e., a binary classifier that predicts if a given comment is either a violation or not, as marked by the moderators).
We introduce additional supervision into our dataset of removed comments, by automatically categorizing content based on the types of violations they represent in a large-scale, computational manner.
By adopting a hybrid approach---classification, topic modeling, and open-coding---we map over 40,000 removed comments collected from 100 different communities on Reddit, to the specific type of norms they violate.

\begin{figure*}[!t]
\centering
  \includegraphics[width = 0.8\linewidth]
{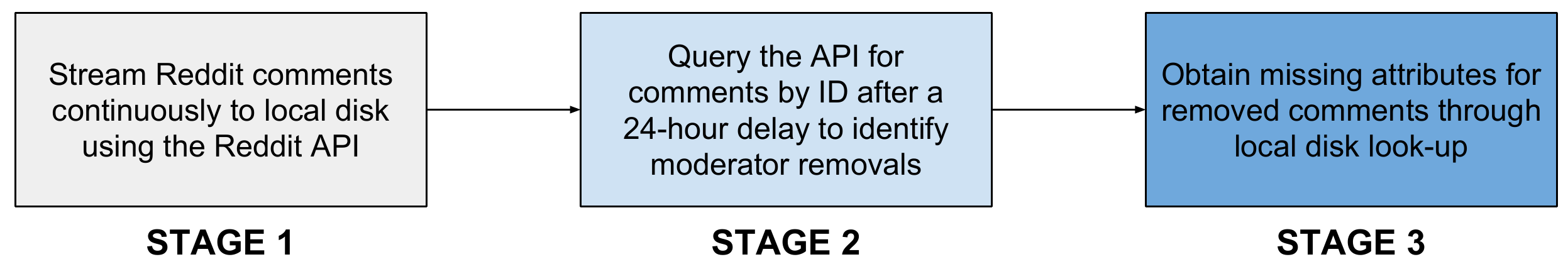}
  \caption{Flowchart describing the three stages involved in our collection of Reddit comments removed by moderators.}~\label{fig:data-collection}
  \vspace{5pt}
\end{figure*}

\subsection{Summary of contributions}
First, we present a three-stage process to collect Reddit comments that were removed by moderators for violating community norms, and publicly release over 2M removed comments from top 100 subreddits.
Next, we introduce a hybrid approach to detect 8 different types of \textit{macro norms} on Reddit (i.e., norms that are widely enforced on most parts of Reddit).
Finally, we automatically label over 40,000 Reddit comments removed by moderators according to the specific type of norms being violated. The dataset~\footnote{\url{https://doi.org/10.5281/zenodo.3338698}} containing all 2M removed comments, along with the 40K labeled macro norm violations generated in this paper can be found  \href{https://github.com/ceshwar/reddit-norm-violations}{online}.
All of the removed comments from Reddit, and the norm extraction (using a mix of classification, topic models and manual annotation) were generated as part of our prior work~\cite{chandrasekharan2018internet}.

\subsection{Ethical considerations} 
We recognize that the use of ``deleted data'' (here in the form of comments removed from Reddit) is controversial territory in social computing research. We discussed these issues in detail with 
the Institutional Review Board 
and our colleagues, before conducting this research. In the end, we concluded that examining removed comments provides key insights into how online communities are governed, and those benefits outweighed any downside risks, as long as the risks are mitigated.
For example, we believe that our findings and data may enable new \textit{mixed-initiative} moderation tools for online communities. We actively worked to minimize potential risks by not linking removed comments in our dataset back to their authors or subreddits (who may not want to be immortalized in a research paper next to their norm violation). Moreover, we did not use \emph{comments that were deleted by their authors} in this work, as those felt qualitatively different to everyone with whom we discussed this research. Finally, we only collected public data through Reddit's official API, in an effort to protect Reddit itself from any harm.

\section{Collecting removed Reddit comments}

Next, we describe our three-stage process to collect Reddit comments that were removed by moderators.
An illustration of this approach is shown in Figure~\ref{fig:data-collection}.

\subsection{Corpus of removed Reddit comments ($\dataset{}$)}

We constructed a dataset that included all Reddit comments that were moderated off-site\footnote{Therefore, these comments were no longer publicly visible on the internet at the time of data collection.} during a 10-month period, from May 2016 to March 2017, in a three-stage process.

\subsubsection{Stage 1: Streaming Reddit comments using Reddit API}
We used the Reddit streaming API~\footnote{\url{https://praw.readthedocs.io/en/latest}} 
to continuously crawl all comments as they were posted on Reddit. Essentially, these were all comments posted to r/all, and they can be from any subreddit that is not ``private,'' and chose to post its content to r/all. 
We stored all the comments obtained from the stream in a master log file that was maintained locally.

\subsubsection{Stage 2: Querying the Reddit API after a 24-hour delay}
After a 24-hour delay, we queried the Reddit API for each comment that was collected in the past day. In order to do this, we used each comment's unique \textit{comment\_ID}, which was stored in our master log file during \textit{Stage 1}.
If a comment is removed by a moderator, then the text that was previously present in the comment (represented in the ``body'' field) is replaced with \textbf{[``removed'']}, and all other attributes of the comment, except the ``\textit{comment\_ID}'' (unique identifier), are discarded. Essentially, none of the fields that were previously associated to a comment are available publicly after its removal. 

It is important to distinguish between a comment that was taken down by a moderator from a comment that was taken down by the author themselves. 
Prior work has shown that it is possible to do so---only when moderators or admins remove a comment, its text is replaced by [``removed''], and most comments violating norms are moderated within the first 24 hours of posting on the subreddit~\cite{chandrasekharan2018internet}. 
Using this method, we compiled the unique identifiers, or ``\textit{comment\_IDs}'', of all comments that were removed by moderators in the last day.

\subsubsection{Stage 3: Obtaining missing data for removed comments}
As mentioned earlier, all the attributes of a comment (except its unique identifier, or \textit{comment\_ID}) are discarded following its removal. In order to retrieve these missing fields, we performed a look-back step using the data present in our master log.
For each removed comment we identified in the\textit{ Stage 2}, we performed a look-up in the master log file (compiled in \textit{Stage 1}), using the \textit{comment\_IDs} of the removed comment.
Through this look-up, we obtained all the fields that were previously contained in the removed comments (like `body', `subreddit', `author', and so on) before it was removed by moderators.

Using this 3-stage process, we collected 4,605,947 moderated comments that were removed from Reddit during a 10-month period, from May 2016 to March 2017. All of the removed comments we identified constitute our Reddit moderated comments corpus (denoted by $\dataset{}$ in the remainder of the paper).

\subsection{Discarding spurious comments in $\dataset{}$}

Next, we discarded comments present in $\dataset{}$, based on three different criteria mentioned below:

\begin{itemize}
    \item \textit{Replies by AutoModerator}: We discarded 101,502 comments from $\dataset{}$, that were authored by \textit{AutoModerator}, a moderation bot used to automatically remove content based on predefined rules.\footnote{https://www.reddit.com/r/AutoModerator/} Since these comments were just warnings issued to users following actual removals, we do not consider them as norm violations in our analysis.
    \item \textit{Replies to removed comments}:
    Through conversations with moderators, we learned that sometimes when a comment is removed, all its \textit{replies} are also removed automatically. So, we discarded 1,051,623 comments from $\dataset{}$, which were identified to be replies to removed comments, since they may not have been removed because they violated a norm (rather because they replied to a comment that did).
    \item \textit{Comments from non-English subreddits}: Next, we used \textit{langdetect}~\footnote{https://pypi.python.org/pypi/langdetect} and examined all comments from subreddits to decide whether interactions are predominantly in English or not. As a result, we discarded all comments from any non-English subreddits present in $\dataset{}$, giving us the final dataset for further analysis.
\end{itemize}

\subsection{Identifying study subreddits}
After preprocessing the data by discarding spurious comments, there remained over 3 million moderated comments contained in $\dataset{}$, and they were collected from 41,097 unique subreddits.
In order to build robust machine learning (ML) classifiers, we restricted our analysis only to the subreddits for which we were able to collect a reasonable amount of moderated comments. 
Therefore, we discarded all subreddits that generated fewer than \textit{5,000} moderated comments in $\dataset{}$.

Finally, 2,831,664 moderated comments remained in $\dataset{}$, all originating within the 100 subreddits generating the most removed comments in our corpus.
We call these 100 subreddits our \textit{study subreddits}\footnote{Names of all 100 study subreddits can be found in the \href{https://github.com/ceshwar/reddit-norm-violations/blob/master/data/study-subreddits.csv}{repo}.}, and denote their removed comments by $\matfff{}$  for the rest of this paper.
We made the entire dataset of removed comments obtained from the 100 subreddits post-filtering publicly available.

\subsubsection{Description of study subreddits.}
At the time of our analysis, the study subreddits had an average 5.76 million subscribers, with r/funny having the highest subscriber count (19 million), and r/PurplePillDebate having the lowest subscriber count (16,000).
On average, each subreddit contributes 20,070 moderated comments, with r/The\_Donald contributing the most (184,168) and r/jailbreak contributing the least (5,616) number of removed comments in $\matfff{}$.

\section{Identifying macro norm violations using subreddit classifiers}

\begin{figure*}[!t]
\centering
  \includegraphics[width = \linewidth]
{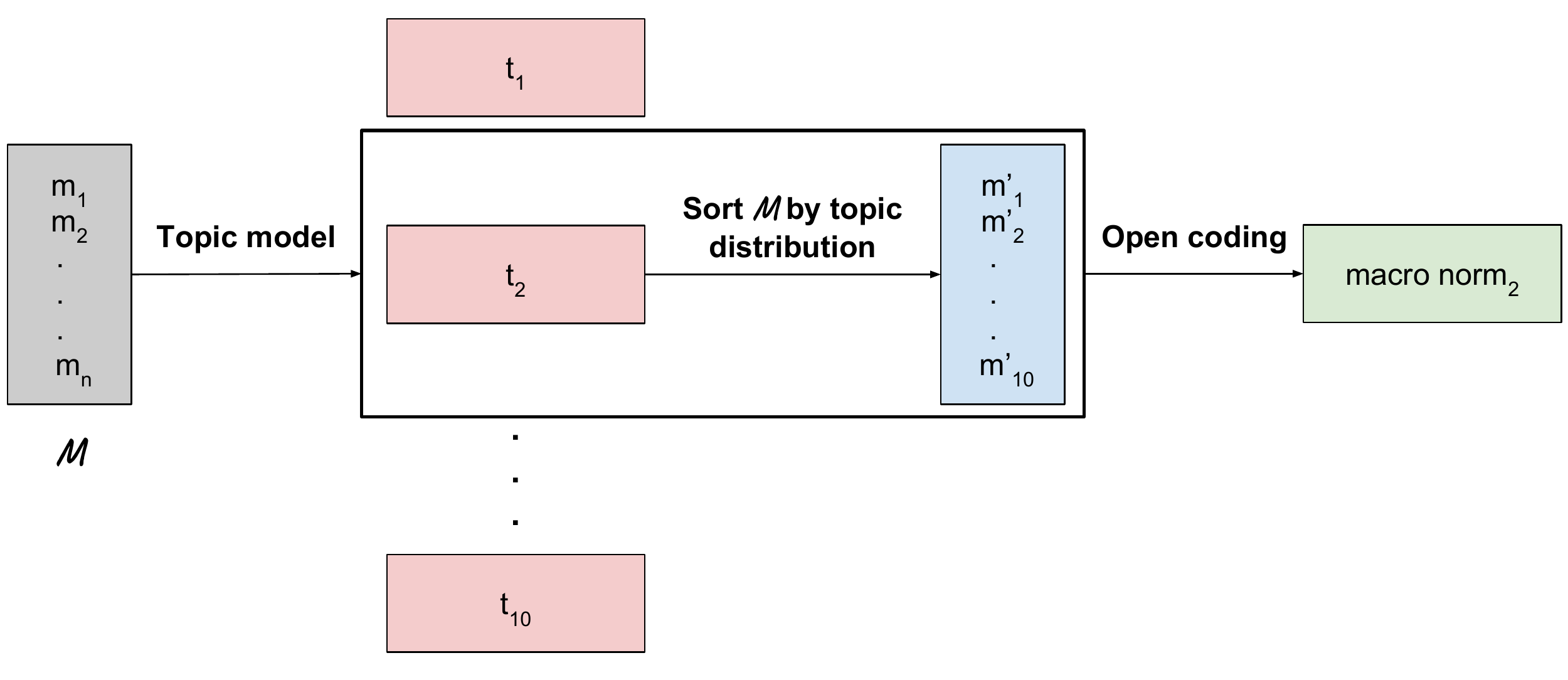}
  \caption{Based on agreement among subreddit classifiers, we identified the comments in $\matfff{}$ that are considered to be norm violations on most parts of Reddit (i.e., macro norm violations). 
  Using these comments, we first performed topic modeling to identify the top 10 topics contained within them.
  Finally, we employed open coding to label the 10 topics we identified earlier, using 10 comments that ranked highly in each of the topics
  as context, by the norms they violated.
Through this hybrid procedure, we extracted \textit{macro norms} that are enforced on most parts of Reddit.}~\label{fig:norm-extraction}
  \vspace{3pt}
\end{figure*}

In our prior work, we have detailed the procedure used to train classifiers that can predict moderator removals within the study subreddits~\cite{chandrasekharan2018internet}.
Using the comments that were removed by moderators of each study subreddit, along with \textit{unremoved} comments collected from study subreddits, we trained machine learning models to predict whether a comment posted on the subreddit will get removed by moderators or not.

\subsection{\textit{Summary:} Training subreddit classifiers}
First, we trained in-domain classifiers to predict whether a comment posted on a subreddit will get removed by a moderator or not. We used \textit{fasttext}, which is a state-of-the-art library for text classification~\cite{bojanowski2016enriching,joulin2017bag}.  
\textit{fasttext} represents each instance by the average of vector representations for words and n-grams, which are short units of adjacent words.

For each study subreddit $S_k$, we built a classifier $clf_{S_k}$ trained on comments removed by moderators of $S_k$, along with an equal number of randomly sampled comments from $S_k$ that were not removed, at the time of our data collection (i.e., \textit{unremoved} comments).
We refer to each in-domain classifier built entirely using removed and unremoved comments from a single subreddit as a ``subreddit classifier''. We built 100 such subreddit classifiers, one for each of our study subreddits.
For detailed descriptions of the construction of our 100 subreddit classifiers, and evaluation of the in-domain classifiers through 10-fold cross-validation tests, please refer to our prior work~\cite{chandrasekharan2018internet}.

\subsection{Agreement among subreddit classifiers}
Next, we obtained predictions from each subreddit classifier (e.g., $clf_{S_k}$) for each comment present in $\matfff{}$, and generated a \textit{prediction matrix}. Columns in this matrix are comments in $\matfff{}$, and rows are subreddit classifiers. Each cell [i,j] in the prediction matrix contains a \textit{yes} or \textit{no}, depending on what classifier $clf_{S_i}$ predicted for comment $m_j$: \textit{If $m_j$ were hypothetically posted on $S_i$, would it get removed by a moderator?}
The output of this step is a \textit{prediction matrix}, with number of rows equal to the number of comments in $\matfff{}$ (2M), and the number of columns equal to the number of subreddits for which we have trained classifiers (100).

Using the prediction matrix, we computed the overall agreement among all subreddit classifiers' predictions for each removed comment present in $\matfff{}$. By overall agreement among subreddit classifiers, we refer to the number of classifiers that predict to remove the same comment.

\subsection{Comments violating macro norms} 
Based on the amount of agreement among subreddit classifiers, we identified comments that a large majority of the 100 subreddit classifiers agreed to remove.
In particular, we identify comments that a large majority (\textit{at least 85 out of 100}) of the study subreddit classifiers agreed to remove.
It is likely that something in the comment violates a norm shared widely on Reddit, and
we consider these comments to be norm violations at a \textit{macro level}, or macro norm violations.
Using the text contained in these comments, we will extract norms that extend across most study subreddits. In turn, we call these \textit{macro norms}, as we observe them to be enforced by moderators of a large majority of our study subreddits.

\section{Extracting macro norms through topic modeling and open coding}
In order to identify the macro norms these comments violated, we employed a combination of topic modeling and open coding on the macro norm violations. 

\begin{table*}[!t]
\renewcommand*{\arraystretch}{1.1}
\centering
\begin{tabular}{l}
\hline
\textbf{Macro norm violation}\\
\hline
Using misogynistic slurs \\
Opposing political views around Donald Trump (depends on originating subreddit)\\
Hate speech that is racist or homophobic\\
Verbal attacks on Reddit or specific subreddits\\
Posting pornographic links\\
Personal attacks\\
Abusing and criticizing moderators\\
Name-calling or claiming the other person is too sensitive\\
\hline
\end{tabular}
\caption{Macro-norms extracted by analyzing removed comments that at least 85 out of 100 subreddit classifiers predicted to remove from their respective subreddits.}
\label{table:macro-norms}
\end{table*}

\subsubsection{Topic modeling.} First, we adopted a computational approach to reduce the dimensionality of our textual data.
We employed topic modeling on the macro norm violations to identify the underlying topics contained in these comments.
Applying Latent Dirichlet Allocation (LDA)~\cite{blei2003latent}, we estimated topic distributions on the comments found to be macro norm violations. 
We used LDA to estimate the topic distributions among \textit{10} topics. 
Every comment we analyzed is considered to be a document for this analysis.
In further analysis, we tested by increasing the number of topics from 10 to 20 for LDA, but observed that no new types of macro norms emerged. As a result, we only estimated topic distributions among 10 topics.

\subsubsection{Open coding: Map topics to types of norm violation.}
Next, we introduced a qualitative step, where we used open coding to manually label each topic by the norm violation it represents (in the form of a 1-2 line explanation behind the comment's removal). 
Using the topic distribution computed for macro norm violations, we identified 10 comments that ranked highly in each of the 10 topics obtained.
Then, three annotators independently labelled each topic by the norm violation it represents, using the 10 comments ranking highly in that topic as context.
This way, we manually labelled all 10 topics (using 10 randomly sampled comments for context) by their respective macro norms.
Then, the three annotators came together to compare the norms they coded independently, and resolved any disagreements.
Through this process, we extracted the 8 macro norms contained in $\matfff{}$.
The raw annotation data, along with all three labels assigned by the independent coders, are available upon request.

\section{Results}
Working with over 2.8M removed comments collected from 100 communities on Reddit, we identified 8 \textit{macro norms}, which are shown in Table~\ref{table:macro-norms}.
We extracted these macro norms by employing a hybrid approach---classification, topic modeling, and open-coding---on comments identified to be macro norm violations in at least 85 out of the 100 study subreddits. 

\subsection{Macro norms on Reddit}
Hate speech in the form of homophobic and racist slurs are considered as norm violations on most parts of Reddit. 
We also found that name-calling, using misogynistic slurs, graphic verbal attacks, and distributing pornographic material are sanctioned.
Comments presenting opposing political views around Donald Trump, either \textit{for} or \textit{against} depending on originating subreddit, are also removed by moderators. 
Such content could potentially lead to highly polarized comment threads, thereby hijacking ongoing discourse towards unrelated topics.
This indicates that such comments are considered to be norm violations on Reddit because they hurt the process of discussion, and not necessarily because they are universally abhorrent.

Another common macro-norm violation is criticizing and abusing moderators. These are mostly instances where members of the community express their discontent with moderator actions (e.g., removing or promoting certain posts, and the need for transparency in the moderation process).
Sometimes, this discontent goes beyond certain specific subreddits, and we found that users verbally attack Reddit (and its admins) due to a variety of reasons (e.g., policy changes, banning communities, and so on). In such cases, the moderators of subreddits intervened and removed such comments.

\subsection{Map comments in $\matfff{}$ to macro norms}
Using the topics labeled by the type of norm violation they denote, obtained in the previous section and shown in Table~\ref{table:macro-norms}, we
transformed all the removed comments present in $\matfff{}$. 
Essentially, we mapped all removed comments in $\matfff{}$ to the respective macro norms they are likely to have violated.
Using the pre-trained LDA topic models, 
we obtained topic distributions for removed comments in $\matfff{}$. 
This way, we get a sense for the distribution of labeled topics, that were learned by training a topic model on removed comments 
that were considered to be \textit{macro norm violations} (i.e., agreed to be removed by at least 85\% of the study subreddit classifiers).
This approach enabled us to examine and compare the landscape of incoming norm violations, which were eventually removed by subreddit moderators, in a large-scale, computational manner.

The labeled topics, along with their word distributions, and types of norm violations are shown in the Github repo.\footnote{\url{https://github.com/ceshwar/reddit-norm-violations/blob/master/README.md}}
For each of the labeled topics, we identified the \textit{top 5000 comments} in $\matfff{}$ that were \textit{best fit} by the topic model.
In this way, we identified over 5000 removed comments that are examples of each type of norm violation shown in Table~\ref{table:macro-norms}. The removed comments were sorted by their \textit{topic fit}, stored into respective files based on the type of norm violation they represent, and made available on the Github repo\footnote{\url{https://github.com/ceshwar/reddit-norm-violations/tree/master/data/macro-norm-violations}}.

\section{Discussion}

After compiling a corpus of over 2M comments ($\matfff{}$) that were removed by moderators of different communities on Reddit, we categorized each comment by the type of norm it violated, in a large-scale, computational manner.
Using a combination of classification, topic modelling, and human annotation, we presented a novel hybrid approach to detect macro-level norm violations on Reddit.

\subsection{Training nuanced tools for online moderation}
By breaking down our collection of removed comments into more granular violation-types, our dataset can be used to train more nuanced machine learning classifiers for online moderation.
Classifiers trained to detect comments violating the specific types of macro norms we identified in this paper, can detect subtle norm violations, that a simple binary classifier may not be able to identify consistently.
Online moderation is a highly contextual task, that needs to take community norms into consideration, before taking down any content.
Encoding all the norms of a community in the form of simple \textit{removed or not} decisions is not an easy task, and we may require more sophisticated approaches to detecting undesirable content.
We believe that classifiers trained to detect the specific types of norms being violated offer more granularity and information to moderators, when compared to a simple binary classifier trained to predict whether a comment would be removed or not. This is a line of research that we plan to deeply examine in the near future.

\subsection{Classifiers that learn from norms enforced within other communities}
Our findings document the existence of norm violations that are universally removed by moderators of most subreddits. These include comments that contain personal attacks, misogyny, and hate speech in the form of racism and homophobia. 
The presence of these macro norms are in many ways encouraging, as they indicate that engaging in such behavior is considered a norm violation site-wide. 
We would argue that knowing about the presence of such site-wide norms could also help moderators of new and emerging communities shape their regulation policies during the community's formative stages, and feel more confident doing so.

This discovery of widely overlapping norms also suggests that new automated tools for online moderation could find traction in borrowing data from communities which share similar values, or wish to establish similar norms.
By understanding the types of norms that are valued by the target community, researchers could use classifiers trained on the different types of macro norm violations contained in our dataset.
In some cases, the macro norms identified in this work may even be able to serve as sensible defaults for a new online community.

\subsection{Large-scale approach to examine landscape of norm violations}
For existing and established communities, our approach presents a way to understand the landscape of norm violations that are prevalent on their platform, and estimate the types of violations that their current ML models are good at identifying, allowing them to improve existing ML models by focusing on the types of violations that these models are not as good at identifying. In other words, they can focus on achieving targeted gains in performance in the future iterations of the ML models by systematically training on specific types of violations that they want to consciously improve on, or for which the model has not seen enough examples of in the past. Our hope is that this will enable communities to focus on the types of violations that existing models are not as efficient at identifying, rather than just working on small, incremental performance gains on types that they are already good at identifying.

\section{Limitations}
Our findings hinge on the algorithms we used in our methods---the classifiers we trained, and the topic models we employed can play a role in the types of norms we uncovered.
On the other hand, using these algorithms gives us the ability to study site-wide norms holistically in a large-scale empirical manner, which is not possible to do by manual inspection alone.
While we find our results encouraging, they raise a number of questions, challenges and issues.
Here, we reflect on some of the limitations present in our work, that are important to keep in mind when using our dataset. 

\subsection{Noise introduced by the subreddit classifiers}
The mean 10-fold cross-validation F1 score obtained for the 100 study subreddit classifiers was 71.4\%.
This is comparable to the performance achieved in prior work on building purely in-domain classifiers to identify removed comments within an online community~\cite{chandrasekharan2017boc,chancellor2016post}.
But it is important to note that the subreddit classifiers are still imperfect, and could play a role in the types of comments identified to be violating macro norms (for example, it is possible there exist some false positives in our corpus of macro norm violations). 
Nevertheless we are confident that our methods are robust, especially given that all comments were removed by human moderators, and agreement was computed across 100 different subreddit classifiers.

\subsection{Noise introduced by topic modeling}
We employed topic modeling to reduce the dimensionality of our textual data, and identify 10 underlying topics contained in the comments violating macro norms.
After open-coding these 10 topics, we transformed all comments in $\matfff{}$ to identify their best-fit topics, and label them with the corresponding macro norm. 
Though this process allowed us to label large amounts of removed comments in an efficient manner, there are some potential points of failures to keep in mind. Topic modeling is not perfect, and a lot of bias could be introduced by this process. There could exist a comment that ranks highly in multiple topics, leading it to be present in multiple files, even if only one of these topics is the actual best fit for that comment.
As a result, there could exist comments present in the top 5000 for a topic, which do not actually violate the macro norm represented by the topic under consideration.
In order to minimize this risk, we sorted comments based on their topic fit, and saved the top 5000 \textit{best-fit} comments for each labeled topic into our files.

\subsection{Confounding factors} 
Note that we do not know the exact reason behind each moderator removal, and we do not account for differing levels of moderator activity across subreddits.
This lack of context for the removed comments could make interpreting the reasons behind moderator actions a hard task.
Additionally, our current method of data collection does not give us access to removed content in the form of pictures, GIFs or videos. As a result, we were only able to identify community expectations around textual content.

\section{Conclusion}
We present a three-stage process to collect comments that were removed off-site by moderators of Reddit communities for violating community norms.
Through this procedure, we publicly released a large-scale dataset of moderated data that can be studied by future researchers.
Next, we introduced a hybrid approach to detect \textit{macro norms} on Reddit (i.e., norms that are widely enforced on most parts of of the platform) in a large-scale, computational manner.
Finally, we presented a method to automatically label Reddit comments removed by moderators by the specific type of norms they violate.

\bibliographystyle{ACM-Reference-Format}
\bibliography{ms}

\end{document}